\documentclass[structabstract]{aa}  
\usepackage{graphicx}
\usepackage{txfonts}
\usepackage{longtable}
\begin{document}
   \title{Milliarcsecond structure of water maser emission in two young high-mass stellar objects 
          associated with methanol masers
}


   \author{A. Bartkiewicz
           \inst{1}, 
           M. Szymczak
           \inst{1},
           \and
	   H.J. van Langevelde
           \inst{2,3}}

   \institute{Toru\'n Centre for Astronomy, Nicolaus Copernicus
          University, Gagarina 11, 87-100 Toru\'n, Poland\\
          \email{[annan;msz]@astro.uni.torun.pl}
\and      Joint Institute for VLBI in Europe, Postbus 2, 7990 AA
          Dwingeloo, The Netherlands\\
          \email{langevelde@jive.nl}
\and      Sterrewacht Leiden, Leiden University, Postbus 9513, 2300 RA Leiden, The Netherlands
             }

   \date{Received February 9, 2012; accepted March 12, 2012}

 
  \abstract
   {The 22.2~GHz water masers are often associated with the 6.7~GHz methanol masers but
owing to the  
different excitation conditions they likely probe independent spatial and kinematic regions
around the powering young massive star.}
   {We compared the emission of these two maser species on milliarcsecond scales to 
determine in 
which structures the masers arise and to test a disc$-$outflow scenario where the methanol
emission arises in a circumstellar disc while the water emission comes from an outflow.} 
  {We obtained high-angular and spectral resolution 22.2\,GHz water maser observations
of the two sources G31.581$+$00.077 and G33.641$-$00.228 using the EVN.}
{In both objects the water maser spots form complex and filamentary structures of
sizes 18-160 AU. The emission towards the source G31.581$+$00.077 comes from two distinct
regions of which one is related to the methanol maser source of ring-like shape. In both
targets the main axis of methanol distribution is orthogonal 
to the water maser distribution. Most of water masers appear to trace shocks on a working 
surface between an outflow/jet and a dense envelope. Some spots are possibly related to 
the disc-wind interface which is as close as 100-150\,AU to the regions of methanol emission.}
{}

\titlerunning{Milliarcsecond structure of water maser in two HMSFRs}
 
\authorrunning{A. Bartkiewicz, M. Szymczak \and H.J. van Langevelde}

   \maketitle

\section{Introduction}
High-angular resolution observations of high-mass star forming sites in the 6.7\,GHz
methanol maser emission have revealed a diversity of morphological structures from very simple
composed of a few milliarcsecond (mas) wide spots to complex and extended clouds of arcsecond
sizes (Phillips et al.~\cite{p98}; Walsh et al.~\cite{w98}; Minier et al.~\cite{mi00}; 
Dodson et al.~\cite{d04}; Bartkiewicz et al.\,\cite{b09}; Pandian et al.~\cite{p11}).
Linear structures with a velocity gradient are thought to be the signature of
an edge-on disc or torus
around young massive protostars (Norris et al.\,\cite{n98}; Minier et al.\,\cite{mi00}) but there
is evidence that some of them are associated with outflows (De Buizer\,\cite{db03}) or even
a shock propagating through a rotating dense molecular clump (Dodson et al.\,\cite{d04}).
Arc-like or ring-like morphologies of methanol emission seem to prove a disc scenario 
(Bartkiewicz et al.\,\cite{b09}), although most of those structures do not show signs of rotation
but rather inflow or/and outflow dominate (van Langevelde et al.\,\cite{vl10}), it is
then postulated
that methanol masers arise in the shock interface between the large scale accretion and a circumstellar
disc.   

While the 6.7\,GHz methanol maser line is radiatively pumped in warm ($T\sim 150$ K) and dense 
($n\le 10^8$\,cm$^{-3}$) regions (Cragg et al.~\cite{c05}; Sobolev et al.~\cite{s97}), the 22\,GHz water
maser emission is collisionally pumped and probes denser ($n\ge 10^8$\,cm$^{-3}$) and hotter 
($T\sim 400$ K) gas (Elitzur et al. \cite{e89}). The water maser is often excited in strong shocks, 
driven by young low-mass and high-mass (proto)stellar objects, on a working surface between a outflow
and a dense envelope. A number of morphologies of this emission is observed from collimated jet, wide-angle flows,
expanding shells to equatorial flows (Goddi et al.~\cite{g07}; Moscadelli et al.~\cite{m07}, \cite{m11}; 
Sanna et al.~\cite{s10a}, \cite{s10b}; Torrelles et al.\,\cite{tor11}). One or more types of those 
maser structures occur simultaneously in star forming regions on scale sizes of
a few mas to several arcsec
and is possibly related to the geometry of the envelopes or the mechanisms driving the outflows. 

Our recent VLA survey of 22\,GHz water masers in a sample of 31 methanol sources has yielded 
22 detections (Bartkiewicz et al.~\cite{b11}) where both maser species are excited by the same central 
objects. We noted that a distinct group of methanol sources with ring-like structure show either 
no associated water masers at all or water masers that are distributed orthogonally to the major axis 
of the ring. It is argued that the methanol maser structure traces a circumstellar disc/torus 
around a high-mass young stellar object while the water masers originate in
outflows.
As the VLA data are of moderate angular and spectral resolutions of $\sim$1\arcsec\, 
and 0.65\,km\,s$^{-1}$, respectively, we undertook the VLBI observations to examine a disc-outflow
scenario in the two brightest water maser targets, which show the arc-like 
or ring-like morphologies that are characteristic of methanol masers. 

The source G31.581$+$00.077\footnote{The names of two targets follow the Galactic coordinates
of the brightest methanol maser spots derived by Bartkiewicz et
al.\,(\cite{b09}).} has been recognized as a massive young stellar object
based on 6.7\,GHz methanol maser observations (Szymczak et al.\,\cite{sz00}). 
The methanol maser spots are distributed along an arc or ring of 217\,mas size
(Bartkiewicz et al.\,\cite{b09}), which corresponds to $\sim$1200\,AU for 
the assumed near kinematic distance of 5.5~kpc (Reid et al.\,\cite{r09}).
The detection of an infrared source of bolometric luminosity of 3$\times$10$^4$L$_{\sun}$ 
(Urquhart et al.\,\cite{u11}), 22\,GHz H$_2$O maser emission (Bartkiewicz et al.\,\cite{b11}),
1665 and 1667\,MHz weak OH masers (Szymczak \& G\'erard \,\cite{sz04}),
millimeter molecular thermal lines (Szymczak et al.\,\cite{sz07}; Urquhart et
al.\,\cite{u08}), and 5 and 8.4\,GHz continuum emission at location of 31.582$+$00.075 (i.e.,
9\arcsec ~apart from the maser arc/ring) (Urquhart et al.\,\cite{u09}; 
Bartkiewicz et al.\,\cite{b09}) indicates that this is a cluster of recent star formation.

The source G33.641$-$00.228 detected in the 6.7\,GHz methanol line (Szymczak et al.\,\cite{sz00})   
has an arc spot distribution of length 630\,AU (Bartkiewicz et al.\,\cite{b09}) for the assumed
near kinematic distance of 3.8\,kpc (Reid et al.\,\cite{r09}), which seems to be more likely 
than the far kinematic distance (Zhang priv.comm.).
The site also contains water and OH masers (Bartkiewicz et al.\,\cite{b11}; Szymczak \& G\'erard
\cite{sz04}) but no 8.4\,GHz continuum emission was detected with an upper limit of 0.15\,Jy
(Bartkiewicz et al.\,\cite{b09}). The detection of radio recombination lines at about 103\,km\,s$^{-1}$
(Anderson et al.\,\cite{a11}) and a multi-feature $^{13}$CO line spectrum (Urquhart et al.\,\cite{u08})
illustrates the complexity of this molecular cloud and its clustered
star formation. 

\section{Observations and data reduction}
The EVN\footnote{The European VLBI Network is a joint facility
of European, Chinese, South African and other radio astronomy institutes 
funded by their national research councils.} observations of G31.581$+$00.077 and G33.641$-$00.228
with the antennas at Jodrell Bank, Effelsberg, Medicina, Mets\"ahovi, Onsala,
and Yebes, 
were carried out at 22.23508~GHz on 2010 October 30 for 8~h (the project EB047). 
The tracking phase centres were estimated from the VLA survey for the water maser spots that were
located nearest to the methanol emission in each target (Bartkiewicz et al.\,\cite{b11})
at  
$\alpha$=18$^{\rm h}$48$^{\rm m}$41\fs951,
$\delta$=$-$01\degr10'02\farcs578 and $\alpha$=18$^{\rm h}$53$^{\rm
m}$32\fs563, $\delta$=$+$00\degr31'39\farcs130 (J2000), respectively. 
A phase-referencing scheme was applied with a reference source 
J1851$+$0035 (from the VLA calibrator catalogue), using a cycle time 
between the maser and phase-calibrator of 60~s$+$90~s. This strategy 
yielded 2~h on-source times for each target. The 
bandwidth was set to 8\,MHz and data
were correlated with the Mk\,IV Data Processor operated by JIVE with 1024 spectral channels. 
The resulting spectral resolution was 0.1\,km\,s$^{-1}$. 
The velocity was measured with respect to the local standard of rest (LSR).

The data calibration and reduction were carried out with NRAO's Astronomical Image Processing 
System (AIPS) using standard procedures for spectral line observations. 
We used the Effelsberg antenna as a reference. The amplitude was calibrated 
by performing measurements of the system temperature at each telescope and applying
the antenna gain curves. The parallactic angle corrections were subsequently added to
the data. The source 3C454.3 was used as a delay, rate, and bandpass calibrator. 
The phase-calibrator was imaged and a flux density of 190\,mJy was obtained.
The maser data were corrected for all Doppler effects and self-calibrated using 
the brightest and most compact maser spot. Finally, naturally weighted maps of 
spectral channels were created in the velocity range where the emission was seen 
in the scalar-averaged spectrum. For imaging, the resulting synthesized beam was 
1.0$\times$2.4\,mas$^2$ with a position angle (PA) of $-$38\degr, while the pixel 
separation was 0.2\,mas in both coordinates. The rms noise levels (1$\sigma_{\rm rms}$) 
in line-free channels was typically 10\,mJy\,beam$^{-1}$. The weakest detected maser spot had 
a brightness of 94\,mJy\,beam$^{-1}$ that is more than 9$\sigma_{\rm rms}$. 

The positions of water maser spots in all channel maps were determined by fitting
two-dimensional Gaussian models. The formal fitting errors resulting from the 
beamsize/signal-to-noise ratio were smaller than 0.1\,mas. To determine 
the position accuracy of registered maser spots, we need to consider the following factors
(Diamond et al.\,\cite{d03}): 
i) the uncertainty in the phase-reference source position of 
2\,mas\footnote{The GSFC ICRF2 VLBI Source Position Catalog.};  
ii) the antenna positions, where the claimed 
accuracy of $\sim$1\,cm corresponds to 1\,mas in RA and 2\,mas in Dec; 
iii) the phase transfer over the separation between targets and the phase-calibrator:
1\fdg92 (between G31.581$+$00.077 and J1851$+$0035) and 0\fdg45 (between G33.641$-$00.228 
and J1851$+$0035). These caused potential phase-solution transfer errors 
corresponding to 1\,mas in RA and 2\,mas in
Dec for G31.581$+$00.077 and to 0.2\,mas in RA and 0.4\,mas in   
Dec for G33.641$-$00.228, respectively. In total, the absolute position
accuracy (1$\sigma_{\rm pos}$) is 2.5\,mas in RA and 3.5\,mas in Dec for both targets.

\section{Results}
Water maser emission towards both targets was mapped, after successful phase-referencing, 
in the areas of 5\arcsec$\times$5\arcsec\, and the velocity range explored by Bartkiewicz
et al. (\cite{b11}) using the VLA.
The EVN maps of water maser spots (Figs.  \ref{g31} and \ref{g33}) are overlayed on the VLA maps 
and the 6.7\,GHz methanol maser distributions obtained with the EVN (Bartkiewicz et al.\,\cite{b09}). 
The {\it Spitzer} GLIMPSE maps\footnote{http://irsa.ipac.caltech.edu/data/SPITZER/GLIMPSE/} 
of the 4.5~$\mu$m$-$3.6~$\mu$m emission excess are added.
Following the procedure described in Bartkiewicz et al.~(\cite{b09}), the water maser
spots are analysed to identify individual velocity-coherent maser {\it clouds}. 
Their basic parameters, such as positions, $\Delta$RA, $\Delta$Dec, 
LSR velocities, V$_{\rm LSR}$, and intensities of the brightest spot, S$_{\rm p}$, 
of each cloud are listed in Tables~\ref{table:1a} and \ref{table:1b}. 
The water and methanol maser spectra of single clouds are combined in Figures \ref{g31_gauss} and 
\ref{g33_gauss} with overlays of individual Gaussians, if emission was seen
in at least three consecutive channels. A Gaussian analysis of individual clouds is carried
out and the fitted flux amplitude, S$_{\rm fit}$, the full width at half-maximum, FWHM, the projected extent
between the most separated single spot centres within a cloud, 
L$_{\rm proj}$, and the velocity gradient, V$_{\rm grad}$ are listed in Table~\ref{table:2}. 
The lower limit to the brightness temperature, T$_{\rm b}$, of each cloud is also calculated
according to Eq.~9$-$27 of Wrobel \& Walker (\cite{w99}). 
For comparison purposes, the same analysis is done for the 6.7\,GHz methanol maser 
data (Table~\ref{table:2}) obtained in 2007 June (G31.581$+$00.077) or 2003 June
(G33.641$-$00.228) with the EVN. Below we comment in more detail on each source.

\subsection{G31.581$+$00.077}
\begin{figure*}
   \centering
   \includegraphics[scale=0.75]{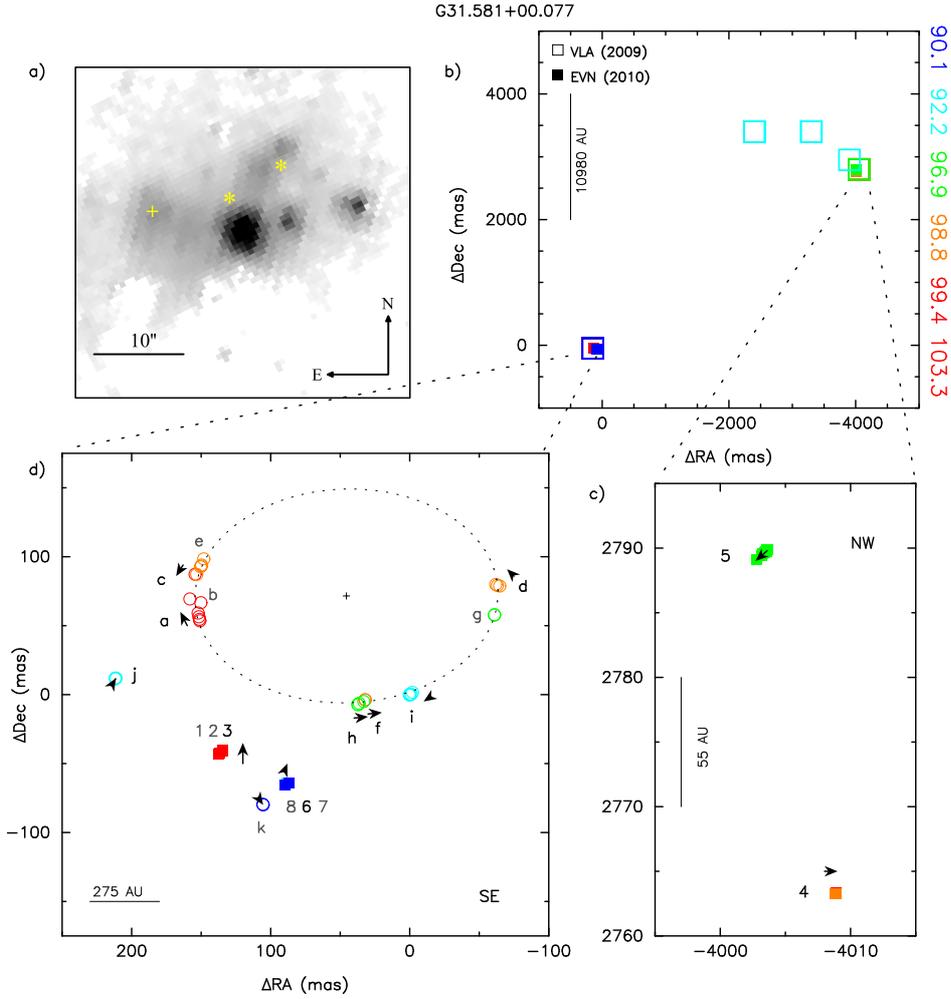}
\caption{G31.581$+$00.077. {\bf a)} {\it Spitzer} GLIMPSE 4.5~$\mu$m$-$3.6~$\mu$m excess image overlaid 
         with the water maser positions (yellow asterisks) from the EVN observation and H~{\small II} region
        (yellow cross) detected by Bartkiewicz et al. (\cite{b09}).
        {\bf b)} Shows the distribution of water masers observed with the EVN (filled squares)
        from this paper and the VLA 
        (open squares) from Bartkiewicz et al.~(\cite{b11}). The colours of
        the symbols relate to the
        LSR velocities as indicated on the right-hand side of the plot. The origin of
        the map is 
        the position of the brightest 6.7\,GHz methanol maser spot (Bartkiewicz et al.\,\cite{b09})
        (see also Table \ref{table:1a}). {\bf c)} Shows an enlargement of the north-western (NW) water masers.
        {\bf d)} Shows an enlargement of the south-eastern (SE) water masers together with the distribution
        of the 6.7\,GHz methanol masers, marked by open circles, from Bartkiewicz et al.~(\cite{b09}). In (c) and
        (d), the
        arrows represent the velocity gradients (from blue- to red-shifted LSR velocities) detected 
        within individual cloud. The black numbers and letters correspond to the clouds with 
        internal velocity gradients, while the grey ones correspond to the clouds without internal 
        velocity gradients (Table~\ref{table:2}).} 
     \label{g31}
    \end{figure*}

A total of 91 water maser spots were detected. They are concentrated in two distinct regions (Fig. \ref{g31}).
The south-east (SE) region of size 50$\times$30\,mas$^2$ containing weak ($<$1.35\,Jy) emission in 
the velocity ranges from 90.1 to 93.9~km~s$^{-1}$ and from 100.0 to 103.3~km~s$^{-1}$ is located
close to the phase centre. 
The second region (NW) of middle velocity emission (96.8$-$100.7\,km\,s$^{-1}$) and flux density 
of 2.6$-$16.5\,Jy is located offset by $\sim$5\arcsec\, to the north-west (PA$=-60\degr$)
of the phase centre.

The SE region is divided into the cluster of clouds {\it 1}, {\it 2}, and {\it 3} separated by 50\,mas from
the cluster of clouds {\it 6}, {\it 7}, and {\it 8} (Fig.~\ref{g31}). 
The clouds in the SE region show Gaussian velocity profiles with FWHM
linewidths of 0.54$-$0.94\,km\,s$^{-1}$. 
Their projected sizes are 0.07$-$0.47\,mas, which correspond to 0.9$-$2.6\,AU for the adopted distance of
5.5\,kpc. The brightness temperature is 
0.24$-$1.38$\times$10$^9$\,K. We note that clouds {\it 3} and {\it 6} show the velocity gradient of 
1.3$-$1.8\,km\,s$^{-1}$mas$^{-1}$ oriented along PAs of 1\degr\, and $-$23\degr, respectively.
It is remarkable that the two clusters of clouds ({\it 1}, {\it 2}, {\it
3} and {\it 6}, {\it 7}, {\it 8}) appear very close in space
(Table~\ref{table:1a}) forming arc-like filaments of projected length of
4 and 3.2~mas corresponding to 22 and 18\,AU, respectively, and 
elongated at PA=$-$58\degr. The closest clouds {\it 3} and {\it 6} are separated by 285\,AU
and a velocity 
difference of 6.7\,km\,s$^{-1}$. The mean linear separation between the two clusters along
PA=65\degr\, is 310$\pm$8\,AU, while the velocity spread is 13.2\,km\,s$^{-1}$.
The velocities of the two filaments are approximately symmetric with 
the regard of the systemic velocity of 96.0\,km\,s$^{-1}$ (Szymczak et al.\,\cite{sz07}).
The two SE filaments are likely signatures of flattened shock surfaces (Torrelles et al.\,\cite{tor01}).
We found that the water cloud {\it 7} coincides within $\pm$0.1\,km\,s$^{-1}$ with the methanol maser cloud {\it
k}, while
the spatial separation is 24.4\,mas which corresponds to the projected distance of 134\,AU.
Although these two lines were observed within a time span of three years
their separation seems to be real.  
This confirms that the water and methanol masers probe different parts of
the environment of young massive
stars (e.g., Beuther et al.\,\cite{be02}; Sanna et al.\,\cite{s10a, s10b}) because of
the different pumping
mechanisms affecting both species (Elitzur et al.\,\cite{e89}; Cragg et al.\,\cite{c05}).    

\begin{figure*}
    \centering
     \includegraphics[scale=0.75]{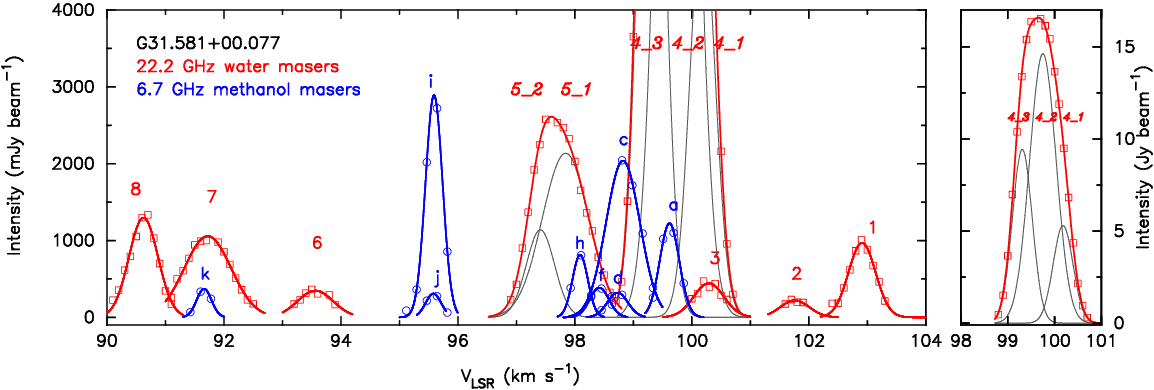}
     \caption{Individual component spectra with Gaussian velocity profiles of 22.2~GHz water masers 
             (red squares and lines) and 6.7\,GHz methanol masers (blue circles and lines) towards 
             G31.581$+$00.077 detected using the EVN in 2010 and 2007, respectively. The numbers and 
             letters correspond to the cloud labels as given in Table~\ref{table:1a} and Fig.~\ref{g31}.
             The grey lines show individual Gaussian profiles fitted to the blended features.
             The clouds with only two single spots are not marked to improve  
             the clarity of the figure (i.e., clouds {\it b}, {\it e}, {\it g}).}
     \label{g31_gauss}
   \end{figure*}

The NW region of elongation of 26\,mas is composed of two clouds {\it 4} and {\it 5} of brightness
temperatures  
of 2.7$-$17.1$\times$10$^9$\,K (Table~\ref{table:2}). 
Their velocity profiles are obviously asymmetric and nicely fitted by the sums 
of three and two Gaussian components (Fig.~\ref{g31_gauss}) of FWHM linewidth of 0.5$-$0.9\,km\,s$^{-1}$. 
The velocities of the NW water maser emission largely coincide with those of
the methanol masers but
their linear separation is about 27000~AU. 
The map of 4.5~$\mu$m$-$~3.6~$\mu$m emission excess (Fig.~\ref{g31}),
which possibly traces shocked molecular gas in outflows from massive stars (e.g., Davies et al.\,\cite{da07}),
shows that the SE and NW maser clusters lie in a very complex region. Both maser clusters are likely
associated with different powering sources.       

The comparison of our maps with those obtained about 14 months earlier with the VLA in
a CnB configuration 
(Bartkiewicz et al.~\cite{b11}) implies that the cloud {\it 1} coincides to within 13~mas, the
cloud {\it 4} to within 56~mas, and the cloud {\it 7} to within 64~mas with the matching VLA spots. 
The emission seen with the VLA from three components located $\sim$1$-$2\arcsec\, eastward 
of the clouds {\it 4} and {\it 5} is not detected. We also note that 
the water maser components detected have very similar amplitudes in both EVN and VLA observations.


\subsection{G33.641$-$00.228}
\begin{figure*}
   \centering
   \includegraphics[scale=0.8]{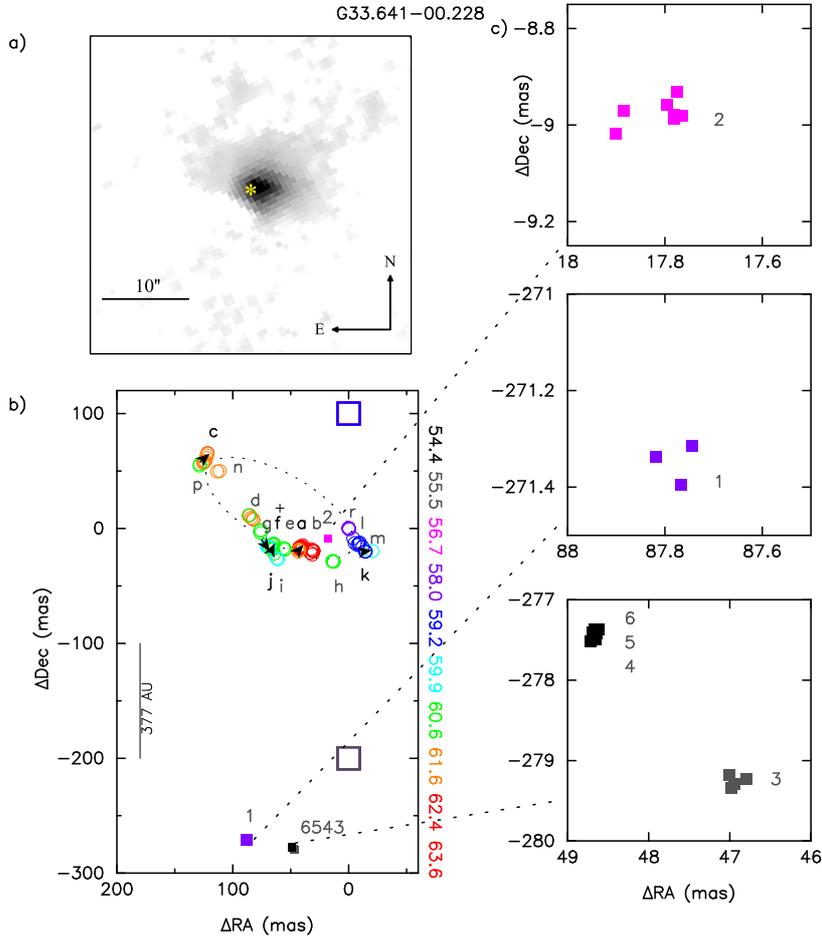}
     \caption{Same as Fig.~\ref{g31} but for G33.641$-$00.228.}
     \label{g33}
   \end{figure*}

Twenty-one water maser spots were detected in the velocity range of 54.4$-$57.6\,km\,s$^{-1}$ over 
the north-south elongated area 50$\times$280~mas$^2$ close to the phase centre (Fig. \ref{g33}). 
Using the above-mentioned procedure, we found four maser clouds where the emission is seen
in at least three contiguous spectral channels. There are also two other clouds ({\it 5} and {\it 6}) 
that do not obey this criterion but seem to be real (Table \ref{table:1b}).
All but one of the cloud are in the southward cluster at a distance $\sim$270~mas from the phase centre.
This filament cluster of size 41.6\,mas is aligned along a PA of 78\degr. For the adopted distance
of 3.8\,kpc, the corresponding linear scales are 1020~AU and 157\,AU. 
The clouds {\it 1}$-${\it 4} in the southern cluster have Gaussian profiles with FWHM linewidths of 
\,0.25$-$0.88\,km\,s$^{-1}$. We note that none of the water clouds have an internal velocity gradient.  
The linear size of the individual clouds is 0.3$-$1.0\,AU and the brightness
temperature is 0.12$-$0.27$\times$10$^9$K (Table \ref{table:2}). 

\begin{figure*}
   \centering
   \includegraphics[scale=0.8]{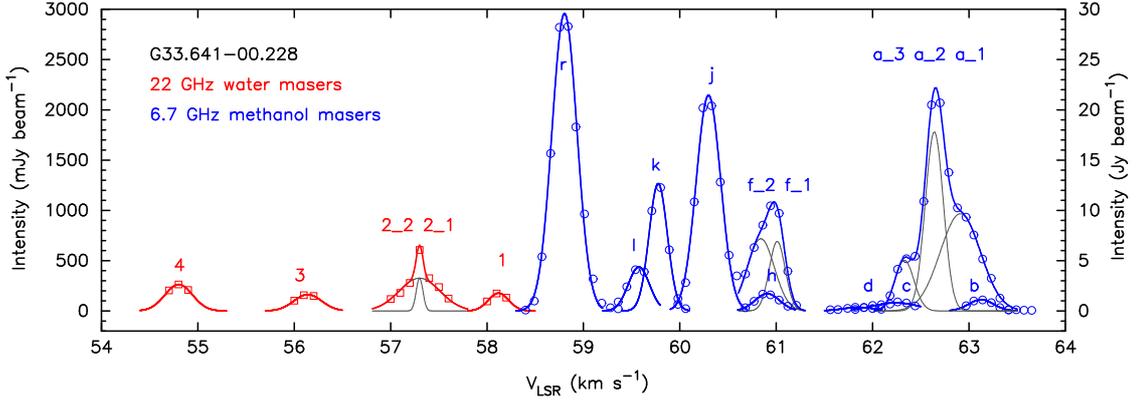}
     \caption{Same as Fig.~\ref{g31_gauss} but for G33.641$-$00.228. The
              scale on the left side describes the water maser intensity, while the
              right side corresponds to the methanol maser intensity. The clouds with only 
              two single spots or without Gaussian characteristic are not marked for the 
              clarity of the figure (i.e., clouds {\it 5}, {\it 6}, {\it e}, {\it
              g}, {\it i}, {\it n}, {\it p}).}
     \label{g33_gauss}
    \end{figure*}

The strongest emission (T$_{\rm b}=0.63\times10^9$K) comes from cloud {\it
2}, which is a blend of 
two Gaussian components (Fig. \ref{g33_gauss}). Cloud {\it 2} is $\sim$20~mas (75\,AU) away
from 
the brightest methanol maser clouds {\it a} and {\it r} and differ in terms
of velocity by 1.5\,km\,s$^{-1}$. 
This is probably the first such tight association of methanol and water masers.
We note that the emission of cloud {\it 2} is blue-shifted by 4.2\,km\,s$^{-1}$ and that of cloud {\it a}
is red-shifted by 1.2\,km\,s$^{-1}$ relative to the systemic velocity of 61.5\,km\,s$^{-1}$
(Szymczak et al.\,\cite{sz07}). It is therefore possible that these clouds signpost a shock front 
in which the water emission originates behind the front, while the methanol emission (cloud {\it a}) 
appears outside of the shock interface in the infalling gas.  
 
The image of the 4.5~$\mu$m$-$3.6~$\mu$m emission excess (Fig.~\ref{g33}) shows that the water maser
emission is located 1\farcs2 to the south-east of single mid-infrared object. This is likely 
the source powering the outflow along a PA of 165\degr\, traced by the water masers.   
      
The positions of the clouds {\it 2}, {\it 3}, {\it 4} and {\it 6} differ by about 92\,mas from the VLA
positions of the matching clouds (Bartkiewicz et al.~\cite{b11}), which is well within 
the $\sim$150\,mas 
accuracy of the CnB configuration VLA data.  A weak ($<$0.9\,Jy) red-shifted emission in the velocity range of 
83.8$-$85.1\,km\,s$^{-1}$ seen with the VLA towards cloud {\it 2} (Fig.~\ref{g31}) was not recovered 
in the VLBI observation. The brightness of the emission from cloud {\it 2} determined in the EVN 
observation is only slightly lower than that measured with the VLA. In contrast, the maser spots 
from the southern region (clouds {\it 1}, {\it 3}$-${\it 6}) are about ten times weaker in the EVN maps
that may suggest the presence of diffuse and/or highly variable emission.


\begin{table}
\caption{List of 22.2~GHz water maser clouds as found in EVN observations 
towards G31.581$+$00.077. The
(0,0) point corresponds to the position of the brightest {\it methanol} maser in this source 
(Bartkiewicz et al.~\cite{b09}): RA=18$^{\rm h}$48$^{\rm m}$41\fs94108,
Dec=$-$01\degr10\arcmin02\farcs5281 (J2000).} 
\label{table:1a}       
\centering            
\begin{tabular}{c c c c c c}
\hline\hline  
Cloud & V$_{\rm LSR}$ & $\Delta$RA & $\Delta$Dec & S$_{\rm p}$ & Number\\
 & (km\,s$^{-1}$) & (mas) & (mas) & (mJy\,beam$^{-1}$) & of spots\\
\hline
1 & 102.9& 137.78 & $-$43.02 & 1008 & 10 \\
2 & 101.7& 136.79 & $-$42.58 & 233 & 4\\
3 & 100.2& 134.57 & $-$40.69 & 472 & 7\\
4 & 99.7& $-$4008.86 & 2763.28 & 16534& 20\\
5 & 97.5& $-$4003.61 & 2789.88 &  2577 & 18\\
6 &  93.5&  88.23 & $-$64.04 & 363 & 7\\
7 & 91.7& 86.79 & $-$63.97 & 1000& 14 \\
8 & 90.7 & 89.65 & $-$65.48 & 1333 & 11\\ 
\hline
\end{tabular}
\end{table}  

\begin{table}
\caption{List of 22.2\,GHz water maser clouds as observed using EVN towards G33.641$-$00.228.
The (0,0) point correspond to the brightest {\it methanol} maser in this source
(Bartkiewicz et al.~\cite{b09}): RA=18$^{\rm h}$53$^{\rm m}$32\fs563,
Dec=$+$00\degr31\arcmin39\farcs180 (J2000).}
\label{table:1b}       
\centering            
\begin{tabular}{c c c c c c}
\hline\hline  
Cloud &V$_{\rm LSR}$ & $\Delta$RA & $\Delta$Dec & S$_{\rm p}$ & Number\\
&(km\,s$^{-1}$ & (mas) & (mas) & (mJy\,beam$^{-1}$ & of spots\\
\hline
1 & 58.1& 87.77 & $-$271.39 & 174 & 3\\
2 & 57.3& 17.79 & -8.96 & 606 & 7\\
3 & 56.1& 46.98 & -279.33 & 157 & 4\\
4 & 54.8 & 48.62 & -277.37 & 263 & 3\\
5 & 55.1& 48.65 & -277.49 & 118 & 2\\
6 & 54.5 & 48.69 & -277.40 & 217 & 2\\
\hline\
\end{tabular}
\end{table}

\begin{table*}
\caption{Parameters of 22.2~GHz water and 6.7~GHz methanol maser clouds.}
\label{table:2}      
\centering          
\begin{tabular}{l c c c c c c c c c }    
\hline\hline       
Cloud &S$_{\rm p}$ & S$_{\rm fit}$ & V$_{\rm fit}$ & FWHM 
& \multicolumn{2}{c}{L$_{\rm proj}$}& \multicolumn{2}{c}{V$_{\rm
grad}$} & T$_{\rm b}$\\
& (mJy~beam$^{-1}$) & (mJy~beam$^{-1}$) & (km~s$^{-1}$) & (km~s$^{-1}$) & 
(mas) & (AU)$^{i}$ & (km~s$^{-1}$~mas$^{-1}$) &
(km~s$^{-1}$~AU$^{-1}$)$^{i}$ & ($\times$10$^8$~K)\\
\hline                    
\\
\multicolumn{2}{l}{\bf G31.581$+$00.077} & \multicolumn{8}{c}{\bf 22.2~GHz water masers}\\
1& 1008  & 969 & 102.9 & 0.54 &  0.17& 0.93 &- &- & 10.4\\
2& 233   & 221& 101.8 & 0.58 &  0.07& 0.38 &- &- & 2.4\\
3& 472   & 444 & 100.3 & 0.61 &  0.47& 2.58 &1.3& 0.24 & 4.9\\
4& 16534 & - & - & - & 0.10& 0.55 &21.2& 3.84 & 170.8\\
{\it 4$_{\_}$1} &- & 5302 & 100.2 & 0.50 & -&-&-&-&-\\  
{\it 4$_{\_}$2} &- & 14630 & 99.8 & 0.67 & -&-&-&-&- \\
{\it 4$_{\_}$3} &- & 9445 & 99.3 & 0.50 & -&-&-&-&- \\
5& 2577  & - & - & - &  1.17& 6.42 &1.8& 0.33 & 26.6\\
{\it 5$_{\_}$1} &- &2137 & 97.8 & 0.93 & -&-&-&-&- \\
{\it 5$_{\_}$2} &- &1140 & 97.4 & 0.57 & -&-&-&-&- \\
6& 363   & 347 & 93.6 & 0.72 &  0.35& 1.92 &2.0& 0.36 & 3.8\\
7& 1000  & 1057 & 91.7 & 0.94 &  0.36& 1.98 &- &- & 10.3\\
8& 1333  & 1297 & 90.6 & 0.58 &  0.24& 1.32 &- &- & 13.8\\
\multicolumn{2}{l}{}&\multicolumn{8}{c}{\bf 6.7~GHz methanol masers}\\
a & 1099 & 1223 & 99.6 & 0.39 &  5.8 & 31.84 &0.09 & 0.016 & 3.4\\
b$^*$ & 358 &- &- &- &- &- &- &- & 1.1\\
c & 2045 & 2039 & 98.8 & 0.68 &  7.4 & 40.63 &0.07 & 0.013& 6.4\\
d & 292  & 321  & 98.7 & 0.41 &  2.9 & 15.92 &0.12 & 0.022& 0.9\\
e$^*$ & 493 & -& -&- &- &- &- &- & 1.5\\
f & 376  & 386  & 98.4 & 0.39 &  1.6 & 8.78 &0.23 & 0.042 & 1.2\\
g$^*$ & 349 & -& -& -&- &- &- &- & 1.1\\
h & 809  & 814  & 98.1 & 0.31 &  2.0 & 10.98 &0.18 & 0.033& 2.5\\ 
i & 2722 & 2891 & 95.6 & 0.35 &  2.8 & 15.37 &0.25 & 0.046& 8.5\\
j & 276 & 308 & 95.6 & 0.31 & 0.7 & 4 & 0.72 & 0.088 & 0.9\\
k & 333 & 371 & 91.7 & 0.30 & 3.0 & 16.3 & 0.12 & 0.022 & 1.0\\
\\
\hline
\\
\multicolumn{2}{l}{\bf G33.641$-$00.228}& \multicolumn{8}{c}{\bf 22.2~GHz water masers}\\
1& 174 & 177 & 58.1&  0.25 &  0.11& 0.42 &-&-& 1.8\\
2& 606 & - &-& - &  0.16& 0.60 &-&-& 6.3\\
{\it 2$_{\_}$1} & -& 326 & 57.3 & 0.50 &-&-&-&-&-\\
{\it 2$_{\_}$2} & -& 328 & 57.3 & 0.08 &-&-&-&-&-\\
3& 157 & 158 & 56.3&  0.88 &  0.27& 1.02 &-&-& 1.6\\
4& 263 & 263 & 54.8&  0.34 &  0.08& 0.30 &-&-& 2.7\\
5$^*$& 118 &- &- &- &- &- &- &- &  1.2\\
6$^*$& 217 &- &- &- &- &- &- &- &  2.2\\
\multicolumn{2}{l}{}&\multicolumn{8}{c}{\bf 6.7~GHz methanol masers}\\
a & 20690 & - & - & - &  6.7 & 25.26 &0.20 & 0.053 & 64.8\\
{\it a$_{\_}$1}&- & 9670 & 62.9 & 0.49 & -&-&-&-&-\\
{\it a$_{\_}$2}&- & 17826 & 62.6  & 0.22 & -&-&-&-&-\\
{\it a$_{\_}$3}&- & 5038 & 62.3 & 0.23 & -&-&-&-&-\\
b & 1090  & 1124 & 63.1 & 0.3 &  3.3 & 12.44 &- &- & 3.4\\
c & 893   & 811  & 62.2 & 0.6 &  9.0 & 33.93 &0.07 & 0.019& 2.8\\
d & 365   & 358  & 61.9 & 0.4 &  5.8 & 21.87 &- &-& 1.1\\
e$^{**}$ & 207   &- &- &- & 1.3 &  7.92 &- &-& 0.6\\
f & 10447 & -& -& - & 7.5 & 28.28 &0.07& 0.019& 32.7\\
{\it f$_{\_}$1}& -& 6913 & 61.0 & 0.19 & -&-&-&-&-\\
{\it f$_{\_}$2}& -& 7184 & 60.8 & 0.33 & -&-&-&-&-\\
g$^{**}$ & 356   &- &- &- & 0.5 &  10.56 &- &-& 1.1\\
h & 1672  & 1716 & 60.9 & 0.30 &  1.2 & 4.52&- &-& 5.2\\
i$^{**}$ & 2280  &- &- &- & 3.3 &  22.9& - & - & 7.1\\
j & 20402 & 21448& 60.3 & 0.30 &  2.9 & 10.93 &0.21& 0.056& 63.9\\
k & 12260 & 12665& 59.8 & 0.22 &  6.0 & 22.62 &0.06 & 0.016& 38.4\\
l & 4105 & 4362 & 59.6 & 0.26 & 10.9 & 41.10 &- &-& 12.9\\
n$^*$ & 109 & - &- &- &- &- &- &- & 0.3\\
p$^*$ & 176 & - &- &- &- &- &- &- & 0.6\\
r & 28300 & 29569& 58.8 & 0.30 &  1.2 &4.52 &- &-& 88.6\\
\hline
\multicolumn{10}{l}{$^*$ The emission only in two channels.}\\                  
\multicolumn{10}{l}{$^{**}$ More than two spots in a cloud, but no Gaussian characteristic of
its spectrum.}\\                  
\multicolumn{10}{l}{$^{i}$ Assuming the near kinematic distances, 5.5~kpc
and 3.8~kpc for G31.581$+$00.077 and G33.641$-$00.228, respectively (Sect.~3).}\\
\end{tabular}
\end{table*}

\section{Discussion}
\subsection{Maser environments}
The {\it Spitzer} GLIMPSE map (Fig. \ref{g31}) clearly indicates that the source G31.581$+$00.077
lies in a large (35\arcsec$\times$25\arcsec) complex area with an 4.5~$\mu$m emission excess
in the form of clumped and diffuse structures. This extended emission is
proposed to be a tracer
of shocked gas from regions where outflowing gas interacts with the surrounding medium
(Davis et al.\,\cite{da07}; Cyganowski et al.\,\cite{c09}). The $^{13}$CO(1$-$0) spectrum
obtained with a 46\arcsec\, beam (Urquhart et al.\,\cite{u08}) towards the position 31.5808+0.0757 
has two Gaussian components centred at 96.2 and 109.8\,km\,s$^{-1}$. A weak emission
line is clearly 
detected in the range of 85 to 105\,km\,s$^{-1}$, i.e., in the wings of the first component suggests
the presence of outflows. A signature of ordered motions also appears in the HCO$^+$(1$-$0) 
spectrum (Szymczak et al.\,\cite{sz07}), where in this slightly asymmetric profile the strongest emission
is blueward of the source systemic velocity and can be interpreted as the
result of infall 
(e.g., Fuller et al.\,\cite{f05}). We then find 9\farcs2 east of the SE water maser an
H~{\small II} region of flux
density 10.3 and 15\,mJy at 5\,GHz (White et al.\,\cite{w05}) and 8.4\,GHz (Bartkiewicz et al.\,\cite{b09}), 
respectively. This continuum source coincides to within 0\farcs6 with a clump of 3.6~$\mu$m$-$4.5~$\mu$m emission
excess (Fig. \ref{g31}). The NW water masers are located at the borders of two clumps of excited
gas, and their powering source is unclear although it seems unlikely that the SE and NW masers are associated
with the same central object. A comparison of the present EVN maps with those obtained with the VLA
14 months earlier (Bartkiewicz et al.\,\cite{b11}) implies that there is a 
lack of water masers eastward of the NW clouds (Fig. \ref{g31}).
It is likely that these are diffuse and low intensity masers resolved out with the EVN beam.  
The spectrum obtained with a 40\arcsec\, beam sometime between November 2009 and December 2010 
(Urquhart et al.\,\cite{u11}) indicates that the peak flux density of 125\,Jy at 99.7\,km\,s$^{-1}$ is about 
one order of magnitude higher than that measured with the EVN.
We suggest that the water masers of G31.581$+$00.077 are produced by different young stellar
objects in a complex region composed of at least a few high-mass stars well-signposted by the radio continuum
and the methanol and water masers. 

The water maser in G33.641$-$00.228 lies only 1\farcs2 south-eastward of the maximum of 
the 3.6~$\mu$m$-$4.5~$\mu$m emission excess of cometary-like morphology (Fig. \ref{g33}). 
The HCO$^+$(1$-$0) spectrum obtained towards the position 33.648$-$0.224 (Szymczak et al.\,\cite{sz07})
shows a small dip near 61.5\,km\,s$^{-1}$ and a slight redward asymmetry that
is indicative of outflow motions.
The less optically thin H$^{13}$CO$^+$(1$-$0) transition is detected as marginal red-shifted
emission, which is consistent with an 
outflow. In the region where the OH 1665, 1667, and 1720\,MHz masers were detected  
(Szymczak \& G\'erard\,\cite{sz04}) and all of them peak at 60.2\,km\,s$^{-1}$. The strongest 1720\,MHz emission
and a broad (7.4\,km\,s$^{-1}$) 1667\,MHz absorption profile near 56.2\,km\,s$^{-1}$ 
are indicative of 
shock fronts and a continuum background source, respectively. No continuum emission at 8.4\,GHz was 
detected with a 3$\sigma_{\rm rms}$ limit of 0.15\,mJy~beam$^{-1}$ (Bartkiewicz et al.\,\cite{b09}). 
A weak ($\sim$20\,mJy) 8.6\,GHz hydrogen radio recombination-line near 102.9\,km\,s$^{-1}$ detected at 
position G033.645$-$0.227 (Anderson et al.\,\cite{a11}) differs so greatly in
terms of velocity from
the water masers that this may be a chance projection that is unassociated with the methanol and water masers.
We argue that the water masers of G33.641$-$00.228 as well as the associated methanol and hydroxyl masers
are excited by individual high-mass stars.

\subsection{Physical parameters of circumstellar medium}
Lower limits to the brightness temperature of the SE water maser components associated with G31.581$+$00.077 
are always lower than 1.4$\times$10$^9$K, while the measured linewidths range from 0.54 to 0.94\,km\,s$^{-1}$
are quite commonly in the star-forming sources (e.g., Goldreich \& Kwan\,\cite{g74}; Surcis et al.\,\cite{s11a, s11b}).
Since the kinetic temperature of the masing gas might be expected to be about 400\,K (Elitzur et
al.\,\cite{e89}), 
the intrinsic thermal linewidth given by $\Delta v_{\rm FWHM}=2.35482\times \sqrt{k T_k/m}$, where $T_k$ is a
kinetic temperature, $k$ the Boltzmann constant, and $m$ is a molecular mass should be $\sim$1\,km\,s$^{-1}$. This value is
larger than the observed linewidths 
and suggests that the masers are unsaturated. In the same volume of gas where the water maser cloud {\it 7} and
the methanol maser cloud {\it k} appear to coincide (Fig. \ref{g31}), the methanol thermal linewidth would be
0.75\,km\,s$^{-1}$, whereas the measured value is 0.30\,km\,s$^{-1}$. The narrowing of
the line profile is expected
when the maser is unsaturated. Detailed calculations have shown that in one of the strongest
known methanol sources  
NGC7538 about 92\% of the components are unsaturated (Surcis et al.\,\cite{s11b}). In the source
G33.641$-$00.228, 
the observed linewidths of methanol components are generally narrower than in
G31.581$+$00.077, which suggests that the maser is also unsaturated in this region. 

The mas-scale velocity gradients in both lines are observed in G31.581$+$00.077.
We note 
that the velocity gradient at 22\,GHz is about one order of magnitude higher than at 6.7\,GHz 
(Table \ref{table:2}). Extensive discussion of the velocity gradients for methanol masers by
Moscadelli et al. (\cite{m11b}) suggests that there is a kinematical interpretation
of their origin.
We note that the methanol maser gradients in both targets can reflect the ordered motions on
scales of 10$-$60\,AU. Large velocity gradients for water masers in G31.581$+$00.077 suggest that they 
are related to the outflow motions with velocities larger than 50\,km\,s$^{-1}$ (see Sect. 4.3). 

\begin{table}
\caption{Parameters derived by fitting the kinematics of the rotating
and expanding disc model. The signs $+$ and $-$ of the rotation and
expansion velocities refer to the clockwise or anti-clockwise rotation and
outflow or inflow for positive i. Both rotation and flow 
are reversed in the case of negative i. 
For each source, both signs together with the sign of i could be reversed
since our model does not give the directions unambiguously. In the case of
G33.641$-$00.228, two model fits are presented as described in Sect.~4.3.
}
\label{table4}      
\centering          
\begin{tabular}{lrrrrr}    
\hline\hline  
\\
Source & V$_{\rm rot}$ &V$_{\rm exp}$ &V$_{\rm sys}^1$ & i&$\chi_{\rm V}^2$ \\
       & (km\,s$^{-1}$)& (km\,s$^{-1}$)&(km\,s$^{-1}$)& (\degr)& \\
\hline
\\
G31.581$+$00.077 & \\
                 & 0.90 & 1.71 & 98.54 & 77.5 & 134\\
G33.641$-$00.228 &\\
D1                 & 0.84 & $-$1.05& 61.00 & 30.0 & 184 \\
D2                 & 3.72 & 1.74 & 58.40 & $-$30.0 & 222\\
\\
\hline
\multicolumn{6}{l}{$^1$ Systemic velocity.}\\                  
\end{tabular}
\end{table}

\begin{table}
\caption{Parameters derived from fitting the biconical outflow model 
of Moscadelli et al.~(\cite{m00}). In the case of
G31.581$+$00.077, two model fits are presented as described in Sect.~4.3.}
\label{table5}      
\centering          
\begin{tabular}{lrrrrrrr}    
\hline\hline  
\\
Source & X$_{\rm o}$ & Y$_{\rm o}$ &V$_{\rm o}^1$ & PA & $\theta$ & $\Psi$ &
$\chi^2$ \\
       & (mas)& (mas) & (km\,s$^{-1}$)&(\degr)& (\degr)& (\degr)& \\
\hline
\\
G31.581$+$00.077 & \\
O1               &72  & 68 &  $-$77.0 & $-$16 & 15 & 83 & 1.7\\
O2               & 110 & $-$46 & 7.7 & 67 & 17 & 45  & 0.7   \\
G33.641$-$00.228 & 15 & 5   & 15.2 & 16 & 9 & 67 & 0.1  \\
\\
\hline
\multicolumn{8}{l}{$^1$ Constant velocity of a maser spot, $+$ for a direction
radially outward}\\
\multicolumn{8}{l}{from the central star.}\\                  
\end{tabular}
\end{table}     

\subsection{Kinematic models}
The two sources investigated in this study possibly belong to a group of objects where a ring-like
or arc-like methanol maser distribution traces a circumstellar disc/torus around a high-mass 
young stellar object, whereas the water maser distribution is orthogonal to the major axis of 
the methanol structure (Bartkiewicz et al.\,\cite{b11}). Furthermore, these objects are not associated 
with detectable continuum emission at cm wavelengths (Bartkiewicz et al.\,\cite{b09}) 
and may represent an early stage of evolution. The new EVN observations of both sources
have shown that all 
of the individual water maser spots detected previously with the VLA (Bartkiewicz et al.~\cite{b11}) 
unfold into complex and filament structures of sizes 18$-$160\,AU. To examine a disc-outflow
scenario in the two sources, we used a model of a rotating and expanding thin disc 
(Uscanga et al.\,\cite{us08}) for the methanol masers and a model of the outflow 
(Moscadelli et al.\,\cite{m00}) for the water masers. Detailed descriptions of the modelling are
given in the above cited works, and elsewhere for methanol masers in discs 
(Bartkiewicz et al.\,\cite{b09}) and outflows (Bartkiewicz et al.\,\cite{b06}).
We note that in the following we analyse the new water maser data with the methanol maser data
obtained 3$-$7 years earlier (Bartkiewicz et al.\,\cite{b09}). However, our inspection of a few sources published
so far and systematic EVN monitoring demonstrate that the overall methanol maser morphologies are stable 
on a scale time of 6-8~years (Bartkiewicz et al., in prep.). 

In Tables 4 and 5,  
we summarize the best-fit values of both models. For the
rotating and expanding thin disc, the rotation (V$_{\rm rot}$), expansion (V$_{\rm exp}$), and systemic 
(V$_{\rm sys}$) velocities as well as the inclination angle, i, i.e., the angle between the
line-of-sight and the normal to the ring plane defined to be i$={\rm acos} (\frac{\rm b}{\rm a})$ is given
for each source. The solutions were based on the minimization of the $\chi_{\rm V}^2$ function given
by eq. 8 in Uscanga et al. (\cite{us08}). The outflow model for water masers is characterized by 
the vertex of the cone (X$_{\rm o}$, Y$_{\rm o}$), the x--axis coinciding with the projection of the outflow 
on the plane of the sky at the position angle PA, the inclination angle between the outflow axis and 
 the line-of-sight (i.e., the z--axis) $\Psi$, and the opening angle of the outflow/cone 2$\theta$. 
The systemic LSR velocities, V$_{\rm c}$, were assumed to be the same as given in Sect.~3.  
The $\chi^2$ function was assumed to be expressed as  
in Eq. 3 of Moscadelli et al.  (\cite{m00}).

\begin{figure*}
   \centering
   \vspace*{1cm}
   \includegraphics[scale=1]{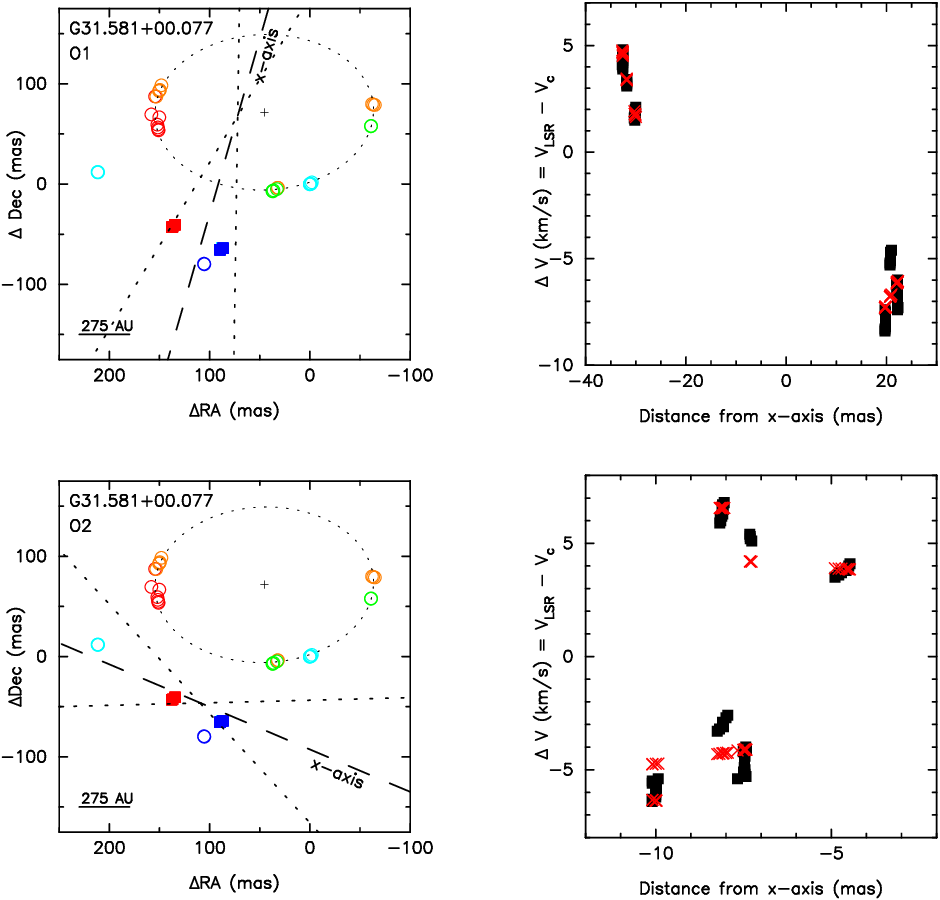}
   \includegraphics[scale=1]{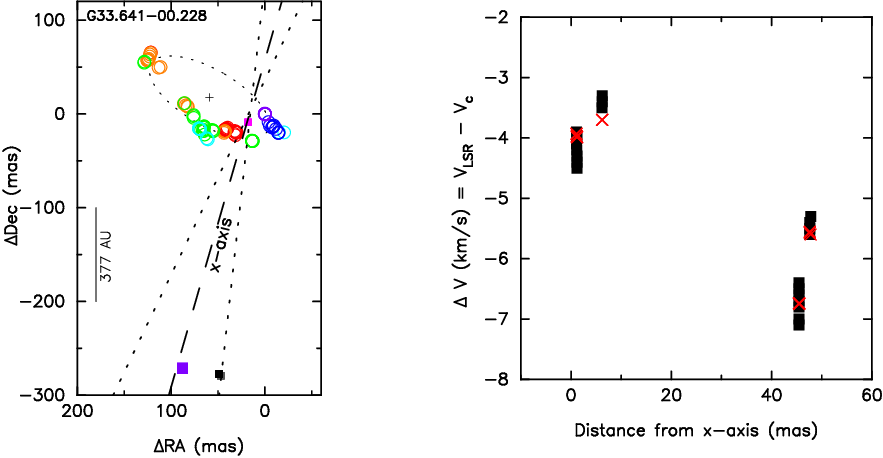}
     \caption{Outflow models fitted to water (squares) 
     in G31.581$+$00.077 and G33.641$+$00.228 according to the
     model of Moscadelli et al.~(\cite{m00}). The relevant parameters are listed in 
     Table 5. The right panel presents a comparison of obtained data (squares) vs. model (crosses). 
     V$_{\rm c}$ is the systemic LSR velocity as given in Sect.~3.}
     \label{models_outflow}
    \end{figure*}

The methanol emission in G31.581$+$00.077 is most accurately reproduced by the model where
the powering source is at the centre of the best-fit ellipse and the rotation velocity
of 0.9\,km\,s$^{-1}$ is a factor of two lower than the expansion velocity (Tab. \ref{table4}, Fig. \ref{models_outdi}).
These values are typical of the class of ring-like methanol masers (Bartkiewicz et al.\,\cite{b09})
and suggest that the methanol structure of $\sim$1000\,AU diameter cannot be interpreted as
a Keplerian disc. It is instead proposed that the methanol masers arise on the interface region
between a large-scale accretion flow and a stellar disc (van Langevelde et al.\,\cite{vl10}; 
Torstensson et al.\,\cite{t11}). We also note that the best-fit 
systemic velocity of 98.54~km~s$^{-1}$ corresponds well to that estimated
from molecular lines (Sect.3.1). 
The SE water emission is reasonably well-fitted by two different models.
In the first model, O1, the vertex of the cone with 2$\theta$=30\degr\, coincides with the centre of
the methanol 
ellipse within 25\,mas and the velocity of masers is 77\,km\,s$^{-1}$ (Tab. \ref{table5}, Fig. \ref{models_outflow}).
We argue that this fit is consistent with the outflow scenario.
 
As we failed to identify any mid-infrared (MIR) counterpart to the centre of
the methanol ellipse as 
a powering source of the methanol structure and perpendicular to the water maser
outflow, we propose 
an alternative model. This second model, O2, assumes that the position of the cone vertex
is just between the red-shifted
and blue-shifted voids of SE water masers. The best-fit velocity of the outflow is
$\sim$8\,km\,s$^{-1}$,  
while the projection of the axis outflow onto the plane of the sky is roughly parallel to the main axis
of the methanol structure (Tab. \ref{table5}, Fig. \ref{models_outflow}). In star forming
regions,  
different centres of activity are reported on scale sizes of a few to several hundred AUs 
(e.g., Torrelles et al.\,\cite{tor01,tor11}). The proper motion studies of G31.581$+$00.077 can
clearly provide conclusive evidence for or against this hypothesis.  

In the case of methanol masers in G33.641$-$00.228, we present the two kinematical models for a powering source 
located at the centre of the best-fit ellipse ($\Delta$RA, $\Delta$Dec)=(59~mas,17~mas) and that located 
at the MIR position ($-$495~mas, 380~mas). In the first model, D1, the infall velocity of about 1\,km\,s$^{-1}$
is slightly higher than the rotation velocity (Tab. \ref{table4}). The structure of water masers is fitted
by a narrow ouflow of 2$\theta$=18\degr\, and velocity of 15\,km\,s$^{-1}$ (Tab. \ref{table5}, 
Fig. \ref{models_outflow}). It is consistent with a jet-like outflow of water masers roughly perpendicular to
the main axis of methanol structure.  
In the second model, D2, the visible methanol structure is only a small part of
the disc radius of $\sim$1900\,AU 
and the rotation velocity of 3.7\,km\,s$^{-1}$ (Tab. \ref{table4}), which implies
that the enclosed mass is about 30M$_{\sun}$. 
In this case, the water masers may trace the wind gas from a disc. 

We note that the kinematic models are unable to accurately reproduce the observed structures of the methanol and water masers
in both targets, and that only measurements of the proper motions of maser spots of the two species can unambiguously
constrain the proposed scenarios.

   \begin{figure}
   \centering
   \vspace*{1cm}
   \includegraphics[scale=1]{model_g31.585.epsi}
   \includegraphics[scale=1]{model_g33.641.epsi}
     \caption{Velocity of the methanol maser spots (open circles) in G31.581$+$00.077
              and G33.641$+$00.228 versus azimuth angle measured from the major axis. 
              The sinusoidal line represents the best-fit kinematical model of a rotating 
              and expanding disc of infinitesimal thickness to the methanol
              spots with the parameters listed in 
              Table 4. For completeness, the water maser spots are shown as squares.
              In a case of G33.641$+$00.228 two kinematical models are
              presented for a powering source located as follow: D1 -- in the centre of the best-fitted
              ellipse, D2 -- at the MIR position.}
     \label{models_outdi}
    \end{figure}

\section{Conclusions}
We have carried out high-angular resolution studies of the 22.2~GHz of water maser line towards 
two methanol maser sources G31.581$+$00.077 and G33.641$-$00.228 using the EVN. 
Although their morphologies did not differ significantly from the previous VLA results, 
the astrometry at the mas level and the properties of the maser clusters could be estimated 
owing to the high-angular and spectral resolution. 
In total, we detected eight and six water maser clouds towards G31.581$+$00.077 and G33.641$-$00.228, 
respectively. In the first target, the water maser components are associated with different centres
of star-forming activity, and the components associated with the methanol ring-like structure 
possibly trace the outflow. In the source G33.641$-$00.228, southern water masers possibly trace
a wind from a disc. The kinematic models containing ring-like or arc-like methanol 
maser structures are able to trace a circumstellar disc/torus around a high-mass young stellar object, whereas 
the water maser distribution is orthogonal to the major axis of the applied methanol structure 
and poorly constrained by the present data. 
The present studies show that the two sources are good targets for proper motion studies in order
to understand more clearly the kinematics of gas in the environments of high-mass stellar objects.
They also encourage us to extend the multi-epoch EVN observations for a whole sample.

\begin{acknowledgements}
AB and MS acknowledge support by the Polish Ministry of Science and
Higher Education through grant N N203 386937. 
This work has also been supported by the European Community
Framework Programme 7, Advanced Radio Astronomy in Europe, grant agreement
nl.: 227290. 
\end{acknowledgements}

\end{document}